\documentclass[10pt,twocolumn,aps,pre,amssymb]{revtex4-2}

\usepackage[utf8]{inputenc}
\usepackage{amsmath}
\usepackage{amssymb}
\usepackage{graphicx}
\usepackage{xcolor}

\usepackage[title]{appendix}

\begin{document}
\title{Ideal Gas Law for a Quantum Particle}

\newcommand{\affCONICET}{Consejo Nacional de Investigaciones
Científicas y Técnicas (CONICET), Godoy Cruz 2290 (C1425FQB), CABA, Argentina}
\newcommand{\affCNEA}{Dto. de F\'{i}sica Te\'{o}rica, GIyA, Comisi\'{o}n Nacional de Energ\'{i}a At\'{o}mica (CNEA), Libertador 8259 (C1429BNP), CABA, Argentina}

\author{Alejandro M.F Rivas}
\author{Eduardo G. Vergini}
\author{Leonardo Ermann}
\author{Gabriel G. Carlo}
\affiliation{\affCNEA} 
\affiliation{\affCONICET} 

\date{\today}

\begin{abstract}
The question of how classical thermodynamic laws emerge from the underlying quantum substrate lies at the foundations of physics. Here, we examine the validity of the ideal gas law (IGL) for a single quantum particle confined within a two-dimensional cavity. By interpreting the quantum wave function as a probability density analogous to that of an ideal gas, we employ the energy equipartition principle to define the temperature of the quantum state. For the mean pressure we take two definitions, one straightforwardly based on the radiation pressure concept and the other taking advantage of a quasi-orthogonality relation valid for billiard eigenstates. We analyze systems with regular dynamics-the circular and rectangular billiards-and compare them with the classically chaotic Bunimovich stadium. We find that the IGL for the first definition of pressure holds exactly in isotropic systems (as the circular case), while for anisotropic geometries, quantum eigenfunctions generally conform to the IGL only on average, exhibiting meaningful deviations. These deviations are diminished in the presence of chaotic dynamics and for coherent states. This observation is consistent with the Eigenstate Thermalization Hypothesis (ETH). Notably, the second definition of pressure allows for a good matching with the IGL.
\end{abstract}

\maketitle

\section{Introduction}

In order to understand the mechanisms leading to ther-
modynamic principles from basic ones we can try to ex-
tend them to the quantum regime \cite{StaffordPRL,Stafford2025}.
 This can provide a bridge between fundamental physics and the most advanced applications of it, such as quantum technology and computation \citep{Gemmer2009, Kosloff2013, Vinjanampathy2016,Tasaki1998}.
This unavoidably touches on fundamental issues of ergodicity and thermalization in isolated systems \cite{Gemmer2009,Tasaki1998} particularly in regimes where coherence and chaos interplay \citep{Gutzwiller1990, Zurek2003}. In particular, the ideal gas law (IGL) in three dimensions \( PV = nRT \) is a cornerstone of classical thermodynamics and a perfect example to illustrate this. 
The IGL  can be derived from kinetic theory under the assumption that particle momentum distributions are ergodic \citep{Boltzmann1896}, ensuring the equivalence of time and ensemble averages and connecting macroscopic variables (pressure \(P\), volume \(V\), and temperature \(T\)) to the microscopic dynamics of non-interacting particles.

Remarkably, recent studies have shown that the IGL persists even for small classical systems, such as few-particle gases in confined cavities \citep{Ciliberto2013,Pathria2011, Urrutia2008}, reinforcing the pivotal role of chaos and ergodicity in sustaining thermodynamic relations connecting microscopic dynamics to macroscopic laws \cite{Arnold1968}.
In classical systems, ergodicity guarantees the equivalence of temporal and ensemble averages, often underpinned by chaotic dynamics \citep{Lichtenberg1992, Gallavotti1996}. However, the quantum realm introduces intrinsic non-ergodic behavior due to wavefunction coherence and quantized energy spectra \citep{Neumann2010, Heller1984}. This motivates a fundamental question: \textit{Does the IGL hold for a single quantum particle, and how do chaotic versus regular dynamics influence its validity?}. 

In this work, we answer this for a single free quantum particle confined in a two-dimensional cavity. We establish a connection between the quantum wave function and the thermodynamic properties of an ideal gas by defining temperature via energy equipartition \citep{Qequipart,Bialas_2019} and pressure through the boundary-normal derivatives of the wavefunction \citep{Qpress,BADER2009} in two ways.
To probe the role of dynamics, we consider two classes of cavity geometries: those with regular dynamics (represented by rectangular and circular billiards) and those exhibiting chaotic dynamics, as in the Bunimovich stadium \cite{Bunimovich1979}.

 Our results demonstrate that for the first definition of pressure, the ideal gas law (IGL) holds \textbf{exactly} in the circular billiard due to its rotational symmetry, while it remains valid \textbf{on average} in rectangular and Bunimovich stadium billiards, though with significant dispersion. This dispersion is related to the anisotropy of the billiard but it is reduced by chaotic dynamics and even substantially diminished for coherent states, the most classical like.  We show that these results align with those obtained using a 'diagonal approximation' where cross-terms between eigenfunctions are suppressed. This approximation is related to the eigenstate thermalization hypothesis (ETH) \citep{Deutsch1991, Srednicki1994,Rigol2008}, suggesting universalities for thermalization in isolated quantum systems. For the second definition of pressure that profits from a quasi-orthogonality relation for billiards, the IGL has a good behaviour.

The remainder of the paper is organized as follows. In Section 2, we review how the IGL is derived using kinetic theory for both regular and chaotic billiards. In Section 3, we extend our analysis to quantum billiards. We begin by defining the temperature through energy equipartition and relate quantum pressure to the normal derivative of the wave function along the boundary. Next, in Section 4 we assess the validity of the IGL for  quantum billiards, in first place for the eigenstates of circular, rectangular, and Bunimovich stadium billiards. We then examine the scenario in which the initial state is a coherent state rather than an eigenstate, highlighting the differences observed between chaotic and regular billiards. We compare our findings with the diagonal approximation and relate these results to the Eigenstate Thermalization Hypothesis (ETH). The final section is devoted to concluding remarks.

\section{Classical billiards}

A classical billiard consists of free particles bouncing elastically on the boundary of a cavity.
 The temperature $T$ relates to the kinetic energy via the energy equipartition principle, 
 involving $k_{B}T/2$ 
for each quadratic degree of freedom, where $k_{B}$ is the Boltzmann's constant. 
For the case a free particle in a 2-D cavity we get: 
\begin{equation}
\frac{p^{2}}{2m}=k_{B}T.
\label{eq:EquipartClas}
\end{equation}
This is the definition of temperature implied by kinetic theory. 
On the other hand, the pressure on the boundary due to collisions  is:
\begin{equation}
P=\frac{F}{L}=\frac{\Delta p}{L\Delta t} ,
\end{equation}
that is, the force $F$ exerted on the boundary is divided by its length $L$. Each elastic collision reverts the component, $p_n$, of the  momentum  normal to the boundary. Hence the momentum change is $\Delta p=2 p_n$ while the time between collisions is $\Delta t$  \cite{reif}. 

\subsection{Circular Billiard}
In a circular billiard of radius $R$ the time   $
\Delta t=L_C/v $ where  $ L_C$  is the length of the chord between two collisions and  $v=p/m$ the speed of the particle. For a momentum vector making an angle $\alpha$ with the normal to the boundary such that \(                p_n=p\cos(\alpha)                 \), the length of chord is \(      L_C=2R\cos(\alpha).          \) We then get, for the force on the boundary
\begin{equation}
F=\frac{2 p_n v}{L_C}=\frac{p^{2}}{mR},
\end{equation}
which divided by the perimeter of the  ($ L = 2 \pi R$) yields the pressure  
\begin{equation}
P=\frac{F}{2\pi R}=\frac{p^{2}}{2m\pi R^{2}}.
\end{equation}
Taking into account the kinetic definition of temperature given in Eq. (\ref{eq:EquipartClas}) we obtain the 2D  
IGL, that is 
\begin{equation}
P S=k_{B}T,
\end{equation}
where $S=\pi R^{2}$ is the surface of  the billiard.  Notice that the Boltzmann constant relates to
the ideal gas constant $R_g$ and the Avogadro number $\mathcal{N}_{A}$ as $k_{B}=R_g/\mathcal{N}_{A}$.
The IGL has been analytically obtained for just one particle bouncing on the walls of the circular billiard without needing any initial condition average. In this particular case,   despite  its regular dynamics, the system is isotropic due to the circular symmetry. In addition, the pressure is homogeneous along the boundary of the billiard.

\subsection{Rectangular Billiard}

For a rectangular billiard with lengths $L_{x}$ and $L_{y}$ . The time between collisions with the same wall is the time it takes for the particle to travel to the opposite wall and back, which is $\Delta t=2 L_{x}/ |v_x|$  and 
$\Delta t=2 L_{y}/ |v_y|$ for the walls of length $L_{y}$ and  $L_{x}$, respectively. Hence on 
boundary of length $L_{y}$ we get 
\begin{equation}
F_{x}=\frac{p_{x}^{2}}{m L_{x}} \quad   \text{and } \quad P=\frac{F_x}{L_{y}}=\frac{p_{x}^{2}}{mL_{x} L_{y}},
\label{eq6}
\end{equation}
while for the walls of length $L_{x}$ we have 
\begin{equation}
F_{y}=\frac{p_{y}^{2}}{mL_{y}}  \quad   \text{and } \quad P=\frac{F_y}{L_{x}}=\frac{p_{y}^{2}}{mL_{x} L_{y}}.
\label{eq7}
\end{equation}
As we can observe, differently from the circular case, the pressure on the walls is not homogeneous as long as $|p_{x}|\neq|p_{y}|$.  The regular dynamics of the rectangular billiard implies conservation of $|p_{x}|$ and  $|p_{y}|$ implying that a single trajectory is not able to  sweep isotropically the momentum space.   Performing an average on initial conditions, all directions become equivalent and isotropy is recovered. For the averaged normal momentum we get 
\begin{equation}
<p_n^{2}>=\frac{1}{2}p^{2}.
\end{equation}
Hence the pressure  defined in equations (\ref{eq6}) and (\ref{eq7}) takes the same value 
\begin{equation}
P = \frac{p^{2}}{2mS},
\end{equation}
 with $S=L_{x}L_{y}$ the surface of the billiard. Finally, using the kinetic definition of temperature of Eq. (\ref{eq:EquipartClas}) we obtain 
\begin{equation}
P S=k_{B} T,
\end{equation}
which is the IGL.  Notice that this relation has been obtained by an average on the ensemble of all possible directions (to recover isotropy), while this not the case for a single trajectory.

\subsection{Bunimovich stadium}

The Bunimovich Stadium Billiard is a rectangle of length $L_s$ with semicircles of radius $R$ at each opposite end. It is a paradigmatic system used to study chaotic dynamics \cite{Arnold1968,Bunimovich1979}. It is  ergodic and mixing, implying that all the available phase space is uniformly explored by a typical single trajectory (excluding periodic orbits and some families of parabolic orbits of null measure).  This property ensures that one single orbit (evolved) is enough to obtain the behavior of the microcanonical ensemble, hence validating the IGL equation.

\section{Quantum billiards}

For the case of a quantum billiard, consider a free particle described by a wave function $\psi(x,y)$ in a 2D cavity that satisfies the Schrodinger's equation 
\[
\hat{H}\left|\psi\right\rangle =\frac{\hat{p}^{2}}{2m}\left|\psi\right\rangle =E\left|\psi\right\rangle.
\]
In coordinates representation the momentum operator is $\hat{p}=-i\hbar\overrightarrow{\nabla}$
and the Schrodinger's equation writes 
\begin{equation}
\frac{-\hbar^{2}}{2m}\nabla^{2}\psi(x,y)=E\psi(x,y).
\label{Schrodinger}
\end{equation}
The wave function $\psi(x,y)$ must satisfy Dirichlet boundary conditions, i.e. it must vanish at $B$  the boundary of the billiard: 
\begin{equation}
\psi(x,y)=0\ \forall(x,y)\:\in\,B.
\end{equation}
The energy $ E $ of a particle in a billiard is the kinetic energy, given in terms of the expectation value the momentum  operators $\hat{p}_{x}$ and $\hat{p}_{y}$ as
\[
E=\langle\psi\mid\hat{H}\mid\psi\rangle =\frac{1}{2m}\left(\langle\hat{p}_{x}^{2}\rangle+\langle\hat{p}_{y}^{2}\rangle\right).
\]

\subsection{Quantum Temperature}

In this section we define a quantum temperature in analogy with the classical concept of temperature described previously. As seen before, the (classical) equipartition of energy implied in
\[
E=\frac{p^{2}}{2m}=k_{B}T
\]
now quantized to a temperature operator such that 
\[
k_{B}\hat{T}=\frac{\hat{p}^{2}}{2m}=\hat{H}.
\]
We can then define in the same sense the temperature density or local temperature as
\begin{equation}
k_{B}T(x,y)=\frac{\hbar^{2}}{2m}\overrightarrow{\nabla}\psi^{*}.\overrightarrow{\nabla}\psi=\frac{\hbar^{2}}{2m}\psi^{*}\nabla^{2}\psi\label{eq:localTemp}.
\end{equation}
This is analogous to the local temperature defined for out of equilibrium systems \cite{ghonge,stafford,Localtemp,Qequipart,lipka}.  Local temperature refers to the temperature of a specific point or region within a system. It is a concept that is used in various fields, including quantum electron systems, density functional theory, and quantum field theory on curved spacetime \cite{stafford,Localtemp,puglisi}. In quantum electron systems, local temperature can be measured using a floating thermoelectric probe, and it is consistent with the laws of thermodynamics when the
probe-system coupling is weak and broadband \cite{eschner,lin}.  In density functional
theory, local temperature is used as a descriptor of molecular reactivity \cite{Guo2021}, and it is determined based on the kinetic energy density of a molecular system \cite{nagy,meair}. In quantum field theory on curved spacetime, a local concept of temperature is introduced to account for the varying temperature distribution in the neighborhood of a geodesic and the failure of a global notion of temperature \cite{Buchholz:2006iv} . The definition of local temperature in electronic systems involves the total electron density and Kohn-Sham orbital densities, and it provides a measure of reactivity towards electron-attracting reagents \cite{nagy}. 

On the other hand, for the whole system the temperature is obtained as the mean value so that,  
\begin{equation}
    k_{B}T=\left\langle \psi\right|\frac{\hat{p}^{2}}{2m}\left|\psi\right\rangle =\frac{\hbar^{2}}{2m}\int\psi^{*}\nabla^{2}\psi dxdy=E.
\label{eq:EKTq}
\end{equation}
This defines the temperature through the quantum equipartition principle \cite{Qequipart}.

Alternative definitions of temperature in quantum systems have recently appeared in the literature. Notably, in the context of open quantum systems, nonequilibrium temperature has been defined as the partial derivative of the von Neumann entropy with respect to internal energy \cite{Alipour2021}. In \cite{lipka}, temperature is treated as an operational concept, inspired by the zeroth law of thermodynamics, effective temperatures are introduced to quantify a quantum system's ability to heat or cool a thermal environment. In a conceptually related approach, temperature is treated as an operator in \cite{ghonge}, where its measurement is intrinsically linked to a collapse of the wavefunction.

\subsection{Quantum Pressure}
In quantum mechanics, the pressure exerted by a particle on the boundary
of a container (or a billiard), or radiation pressure, relates to the 
time derivative of the momentum density in a point $\mathbf r$. Specifically, the
momentum density is defined by
\begin{equation}
\overrightarrow{\pi}(\mathbf r,t)=\Re\bigl[  \psi^* \, \hat p \, \psi \bigr]=
 -\frac{i\hbar}{2}\bigl[\psi^*\overrightarrow{\nabla}\psi - \psi\overrightarrow{\nabla}\psi^*\bigr]\,,
\end{equation}
and by using the Schrodinger equation
\begin{equation}
i\hbar\,\frac{\partial}{\partial t} \psi(\mathbf{r}, t) = \hat{H} \psi(\mathbf{r}, t),
\label{eq:Short}
\end{equation}
we get the Cauchy momentum continuity equation \cite{Lancaster}
\begin{equation}
  \frac{\partial}{\partial t} \overrightarrow{\pi}(\mathbf r,t) +
 \overrightarrow{\nabla} \cdot\mathbf{\Gamma}
  = -\,\rho\,\overrightarrow{\nabla} V ,
  \label{eq:Cauchy}
\end{equation}
with $\rho = |\psi|^2$ and $\mathbf{\Gamma} $ the momentum flux or stress 
tensor whose elements are
\begin{equation}
  \Gamma_{ij}
  = \frac{\hbar^2}{4m}
    \Bigl[\psi^*\,\partial_i\partial_j\psi
          - (\partial_i\psi^*)\,(\partial_j\psi)
         + cc 
         )\Bigr]
  - \delta_{ij}\,\mathcal{L}_{\text{kin}}\,,
\end{equation}
where
\begin{equation}
  \mathcal{L}_{\text{kin}}
  = \frac{\hbar^2}{4m}
    \Bigl[\psi^*\,\nabla^2\psi -
    (\overrightarrow{\nabla}\psi^*)\!\cdot\!(\overrightarrow{\nabla}\psi)
           + cc 
           )\Bigr]
  \;.
\end{equation}
For a free particle in a billiard the potential term is null and the momentum 
flux tensor is symmetric. The \emph{quantum pressure} density $P_i(\mathbf r,t)$ 
 in the $i$ direction (note that anisotropic pressure is admitted) 
is identified with the diagonal element of the stress tensor 
\cite{Lancaster,Sharma} as
\begin{equation}
 P_i(\mathbf r,t)= \Gamma_{ii}= \frac{\hbar^2}{m}\Bigl[ \partial_i\psi^*\partial_i\psi
     - \tfrac{1}{2}|\overrightarrow{\nabla}\psi|^2 \Bigr].
\end{equation}
For a point $(x,y)$ on the boundary $B$ where $\psi(x,y)=0$, the only 
pressure is the normal one, 
\begin{equation}
P(x,y)=\frac{\hbar^{2}}{2m}\frac{\partial\psi^{*}}{\partial_{n}}\frac{\partial\psi}{\partial_{\hat{n}}}=\frac{\hbar^{2}}{2m}\left|\frac{\partial\psi(x,y)}{\partial n}\right|^{2}.
\label{eq:PresQPunt}
\end{equation}

In order to obtain the mean pressure we take an average along the boundary $B$
\begin{equation}
P=\frac{1}{L}\oint _{B}  \frac{\hbar^{2}}{2m}   \left|\frac{\partial\psi}{\partial n}\right|^{2}dl
\label{eq:PresionQ}.
\end{equation}
It is worth noticing that for normalized eigenstates  of any billiard, the following quasi-orthogonality
relationship holds \cite{Barnett2000Deformations}
\begin{equation}
\delta_{i,j}= \int \varphi^{*}_i\varphi_jdxdy\simeq\frac{1}{k_i^2 +k_j^2 }   \oint _{B}\frac{\partial\varphi^{*}_i
}{\partial n}\frac{\partial\varphi_j
}{\partial n}r_n dl,
\label{eq:INTIDENT}
\end{equation}
with $r_n$ the component of $\mathbf{r}$ normal to the boundary and $k_i$ the wave number. The error of this relationship is very
small while $|k_i-k_j| L_c < 1$, with $L_c$ a characteristic length of the billiard; for instance, $L_c=R$ in the
circular billiard. Hence, taken into account the following identities 
\[
1=\frac{1}{L}\oint _{B} dl=  \frac{1}{2 S}  \oint _{B} r_n dl, 
\]
the quasi-orthogonality induces an alternative definition of pressure, $P_2$, by considering a weighted average
along the boundary: 
\begin{equation}
P_2=\frac{1}{2 S}  \oint _{B}  \frac{\hbar^{2}}{2 m}     \left|\frac{\partial\psi}{\partial n}\right|^{2} r_n dl. 
\label{eqPresEDU}
\end{equation}
Then, with this definition of pressure we obtain the IGL
\begin{equation}
P_2 S   \simeq k_B T,
\label{eqIGLEDU}
\end{equation}
which holds exactly for eigenfunctions, and it is a good approximation 
for wave functions satisfying $\epsilon_k L_c < 1$ (localized in the spectrum), with $\epsilon_k$ the wave number dispersion. In contrast, a coherent state with mean wave number $k_m$ satisfies $\epsilon_k L_c \sim \sqrt{k_m  L_c }$,
and consequently the IGL ceases to be valid. Nevertheless, as we discuss in the next section, after a relaxation time, for the pressure $P_2 $ averaged in time IGL is recovered.
It is important to note that both  $P$  and $P_2$  represent spatial averages (or weighted averages) taken over the entire boundary of the billiard. In contrast, pressure measurements in practice are typically performed locally, at specific points. However, by collecting such local measurements along the boundary, one can compute the corresponding averages to obtain the global quantities $P$  and $P_2$. 

\section{ Quantum Ideal Gas Law}

In this section we explore the validity of the IGL for quantum billiards of different geometries, i.e. the regular circular and rectangular cases, and the chaotic case corresponding to the Bunimovich stadium. 

\subsection{Ideal Gas Law for Eigenstates}

First, we will assess the applicability of the Ideal Gas Law (IGL) to the eigenfunctions of the various billiard geometries. Notably, the energy quantization dictated by the Schrodinger equation imposes significant constraints on these eigenfunctions, which, in turn, form the basis for describing the behavior of any wave function.

It is important to highlight that the single-electron framework employed here closely mirrors real nanoscale systems where electrons are confined within well-defined cavities: semiconductor quantum dots, electrostatic ion traps, and superconducting microwave resonators. In each case, strong boundary conditions enforce discrete energy levels and permit strong coupling to the cavity’s electromagnetic modes.
Semiconductor quantum dots act as artificial atoms, confining electrons in all three spatial dimensions—resulting in sharp energy spectra and pronounced light–matter interactions \cite{GarciaDeArquer2021}. Similarly, single electrons in electrostatic ion traps (e.g., Paul or Penning traps) are held by inhomogeneous fields with minimal thermal motion \cite{Leibfried2003}. In superconducting circuit QED architectures, single microwave photons are strongly coupled to qubits within superconducting cavities, enabling coherent vacuum Rabi oscillations \cite{Blais2021}.
This coupling underlies cavity QED phenomena such as the Purcell effect, vacuum-Rabi splitting, and photon blockade. By ignoring thermal fluctuations of the cavity boundary, our treatment isolates intrinsic quantum behavior: temperature, pressure, level statistics, and purely quantum dynamics without extraneous thermal noise.

\subsubsection{IGL for Circular Billiard's Eigenstates}

The eigenfunctions of a circular billiard of radius $R$,  (a quantum particle confined in a circular region) can be found by solving Eq. (\ref{Schrodinger}) in polar coordinates, 
with boundary conditions, $\psi(R,\theta)=0$. In the Appendix we derive explicitly the eigenstates of the circular billiard.  Then, using the definitions of  temperature and pressure of Eq. (\ref{eq:EKTq})  and Eq. (\ref{eq:PresionQ}),  we obtain that for the eigenstates of the circular billiard 
\[
PS=k_{B}T.
\]
That is, the IGL is exact for all the eigenstates of the quantum circular billiard as a manifestation of its isotropic nature. Note that for the circular billiard with the origin of coordinates in the center of the circle ( $r_n=1$ ) and both definitions of pressure  Eq. (\ref{eq:PresionQ})  and  Eq. (\ref{eqPresEDU}) coincide, that is we have $P = P_2$. 

 The fact that the ideal gas law (IGL) holds exactly in the circular 
billiard is a consequence of quantum thermalization arising from the
isotropic geometry of the system.

\subsubsection{IGL for Rectangular Billiard's Eigenstates}

Consider now a quantum particle confined to a rectangular billiard with horizontal and vertical sides of lengths $L_{x}$ and $L_{y}$, respectively. The wave function $\psi(x,y)$ must satisfy Eq. (\ref{Schrodinger}) with boundary conditions
\[
\psi(0,y)=\psi(L_{x},y)=\psi(x,0)=\psi(x,L_{y})=0.
\]
The eigenstates of the system are then given by:
\begin{equation}
    \psi_{n_{x},n_{y}}(x,y)=\frac{2}{\sqrt{L_{x}L_{y}}}\sin\left(\frac{n_{x}\pi x}{L_{x}}\right)\sin\left(\frac{n_{y}\pi y}{L_{y}}\right)
    \label{Eq:EIGREC}
\end{equation}
where $n_{x}$ and $n_{y}$ are positive integers. The energy levels $E_{n_{x},n_{y}}$
are discrete and can be labeled with quantum numbers corresponding to the modes of the system,
\begin{equation}
E_{n_{x},n_{y}}=\frac{\hbar^{2}\pi^{2}}{2m}\left(\frac{n_{x}^{2}}{L_{x}^{2}}+\frac{n_{y}^{2}}{L_{y}^{2}}\right)=k_{B}T\label{eq:=000020KTrect}
\end{equation}
The pressure $P$ at a point on the billiard boundary relates to the mean value of the normal derivative of the wave function as in Eq. (\ref{eq:PresQPunt}).  Applying it to the eigenstates of the rectangular billiard (Eq. (\ref{Eq:EIGREC}))
we get the pressure on its walls as: 
\[
P=\begin{cases}
\frac{\hbar^{2}}{2m} \frac{2n_{x}^{2}\pi^{2}}{L_{x}^{3}L_{y}} & \text{on the vertical walls}\\
\frac{\hbar^{2}}{2m} \frac{2n_{y}^{2}\pi^{2}}{L_{x}L_{y}^{3}} & \text{on the horizontal walls}
\end{cases}.
\]
This pressure  is determined by the quantum numbers $n_{x}$ and $n_{y}$ and is inversely proportional to the respective lengths of the walls, then the pressure is non-homogeneous. 
For the mean pressure on the complete rectangular boundary using Eq. (\ref{eq:PresionQ}) and multiplying by the surface of the billiard $S=L_{x}L_{y}$ we obtain
\begin{equation}
P S=\frac{\hbar^{2}\pi^{2}}{2m\left(L_{x}+L_{y}\right)} \left(\frac{n_{x}^{2}L_{y}}{L_{x}^{2}}+\frac{n_{y}^{2}L_{x}}{L_{y}^{2}}\right). \label{PS-Rec-Exa}
\end{equation}
This expression differs from $k_{B}T$ obtained for Eq. (\ref{eq:=000020KTrect}), hence the IGL is not achieved for a rectangular billiard. 

\begin{figure}[t]
\includegraphics[width=0.47\textwidth]{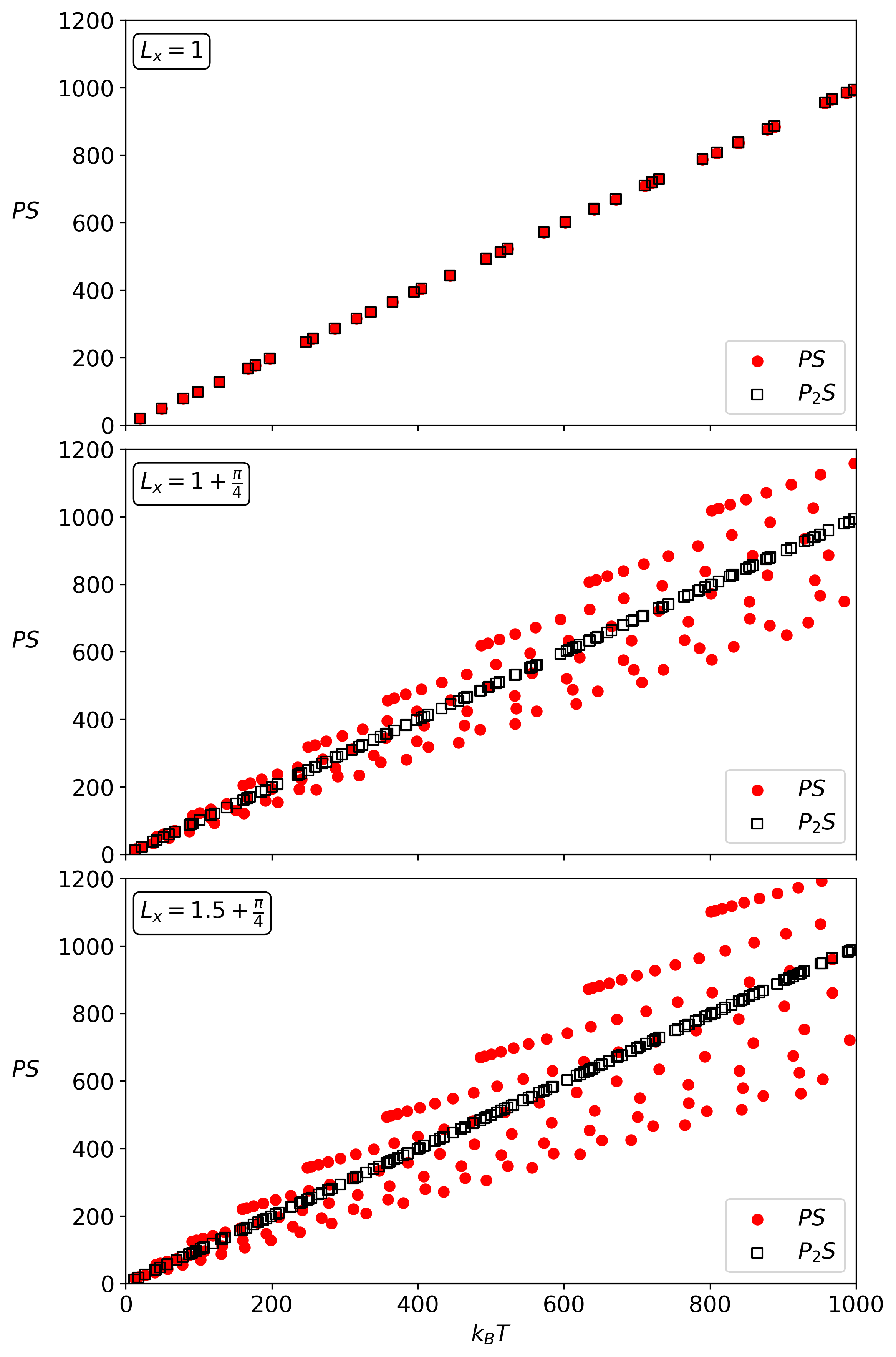}
\caption{$PS$ versus $k_B T$ for eigenstates of the rectangular billiard ($L_y = 1$). The panels display results for different aspect ratios: top ($L_x = 1$), middle ($L_x = 1 + \pi/4$), and bottom ($L_x = 1.5 + \pi/4$). In all panels, filled red circles correspond to $PS$ and open black squares correspond to $P_2S$.}
\label{FigPSvsKT-EIG-REC}
\end{figure}

In Fig. \ref{FigPSvsKT-EIG-REC} we display $PS$ (as in Eq. (\ref{PS-Rec-Exa})) vs $k_bT$ for the eigenstates of the rectangular billiard.  We also display the results for $P_2$ (see Eq. (\ref{eqPresEDU})). 
For all cases, we have taken, $ L_y = 1 $ and different values of $L_x$  ($1,1+\pi/4 $ and $1.5+\pi /4$, in ascending order in the anisotropy). The IGL is not exactly satisfied by all eigenstates; however, they follow it on average. Nonetheless, the maximum difference between the two quantities widens as the temperature increases. Additionally, the dispersion around the IGL increases with the length $L_x$ , i.e. with the anisotropy.  Indeed, for the isotropic case $L_x=1$ the IGL behavior is exactly recovered. Finally, for the alternate definition of pressure $P_2$  given in Eq. (\ref{eqPresEDU}), the IGL is exactly obeyed by all eigenstates.

\subsubsection{IGL for Bunimovich Stadium's Eigenstates}

\begin{figure}
\includegraphics[width=0.47\textwidth]{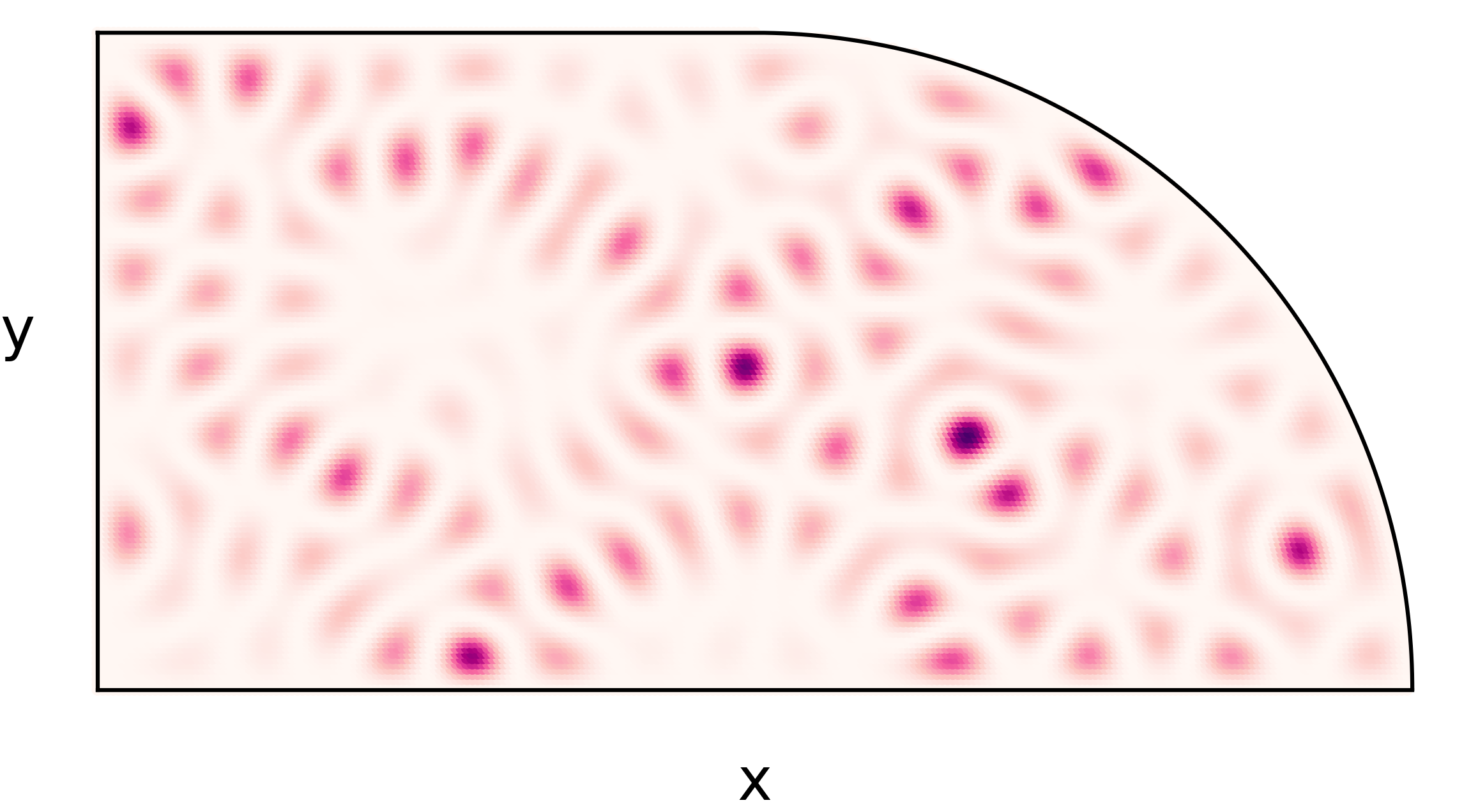}
\caption{Local Temperature density for the $181^{\text{th}}$ eigenstate of the stadium with $l_s=1$ (a typical state) in coordinate space $(x,y)$.}
\label{FigTemp}
\end{figure}

Eigenfunctions of this billiard satisfy Eq. (\ref{Schrodinger}) with Dirichlet boundary conditions. We obtain the eigenfunctions and eigen-energies \cite{EDU1} and calculate the temperature  using Eq. (\ref{eq:EKTq}). 
We perform our calculations on a quarter of the stadium, to avoid symmetry degenerations, a rectangle of length $L_s$ with a quarter of circle of radius $R$ at one of its ends.
Notice that as  we already mentioned, the local temperature Eq. (\ref{eq:localTemp}) is not homogeneous
inside the billiard. In Fig. \ref{FigTemp} we show the local temperature for a typical eigenstate of the stadium.

The pressure on the boundary of the billiard exerted by the eigenvalues is calculated using Eq. (\ref{eq:PresionQ}). 

\begin{figure}[t]
\includegraphics[width=0.47\textwidth]{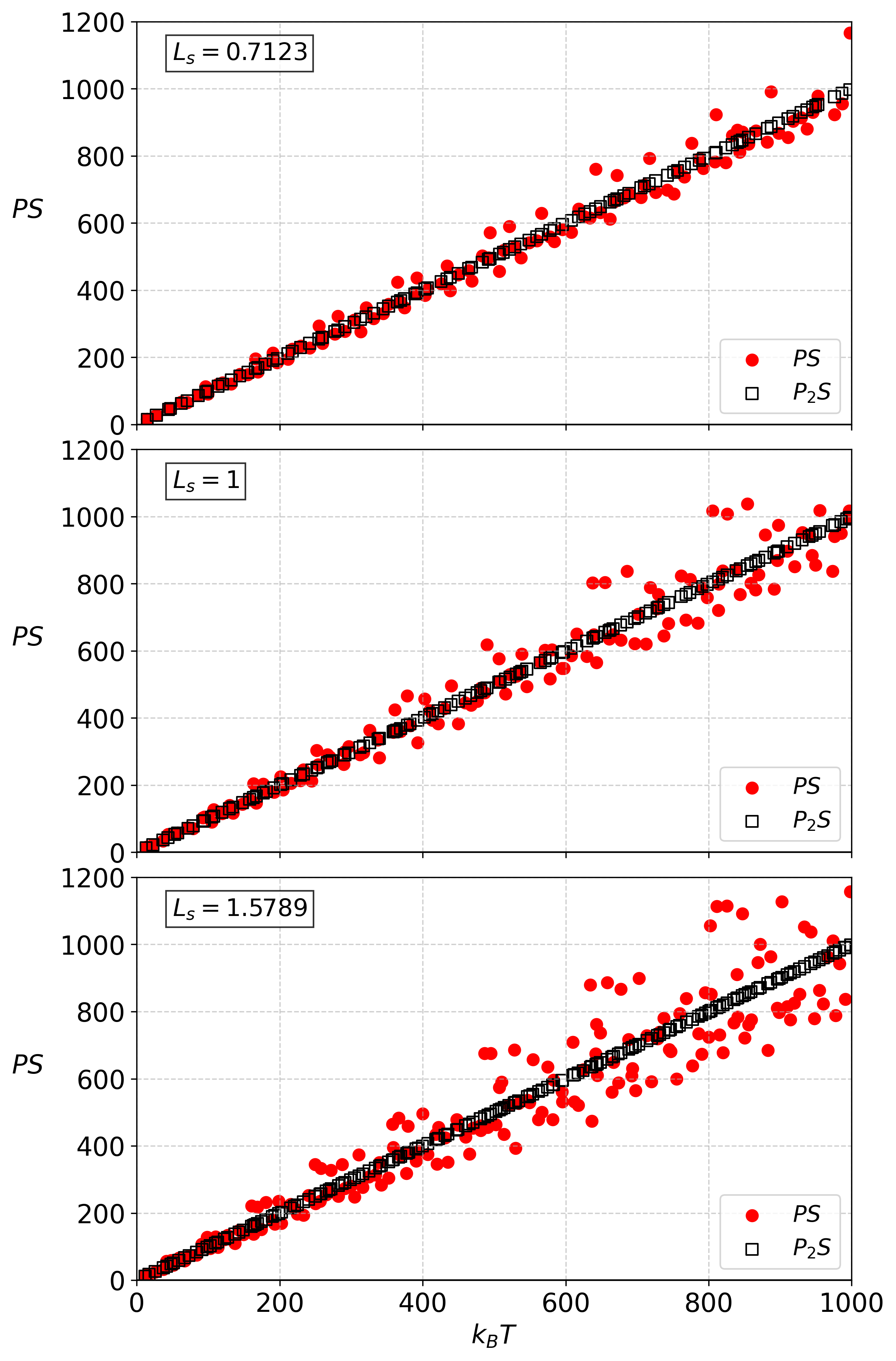}
\caption{$PS$ versus $k_B T$ for eigenstates of the Bunimovich stadium. The panels display results for different aspect ratios: top ($L_s = 0.7123$), middle ($L_s = 1$), and bottom ($L_s = 1.5789$). In all panels, filled red circles correspond to $PS$ and open black squares correspond to $P_2S$.}
\label{FigPSvsKT-EIG-Stad}
\end{figure}

In Fig. \ref{FigPSvsKT-EIG-Stad}  we display for the eigenstates $PS$, the pressure times the stadium surface $S=R(L_{s}+\frac{\pi}{4}R)$ as a function of the temperature $k_{B}T$. We also show the $P_2S$ values.
 For all cases, we have taken, $ R = 1 $ and different values of $L_s$  ($0.7123, 1 $ and $1.5789$, in ascending order in the anisotropy). Although the ideal gas relation is not exactly achieved by all the eigentates, we can say that on average it is, but the maximal difference between both magnitudes grows with the temperature. For $P_2$ the result is exact. 
Also, the dispersion around the IGL increases with the length $L_s$, that is with the anisotropy.   
Notice that, for the isotropic case $L_s=0$ the IGL is exactly achieved since we recover the circular billiard case.
Comparing Fig. \ref{FigPSvsKT-EIG-Stad} with the one for the rectangular case (see Fig. \ref{FigPSvsKT-EIG-REC}) we see that there is a similar behavior but the dispersion around the IGL is lower for the stadium than for a rectangular billiard with the same area (panels b in both Figures).  

\subsection{IGL for Coherent States }

We now explore the case where the initial state is not an eigenfunction of the system but a coherent state (CS) in position $(Q_{x},Q_{y})$ and momentum $(P_{x},P_{y})$
\[
\psi_{Q,P}(x,y)=\left(\frac{1}{\pi\hslash}\right)^{\frac{1}{2}}e^{\left[-\frac{\left(x-Q_{x}\right)^{2}+\left(y-Q_{y}\right)^{2}}{2\hslash}+\frac{i\left(xP_{x}+yP_{y}\right)}{\hslash}\right]}.
\]

\begin{figure}[t]
{\hspace{-1.cm}\includegraphics[width=0.54\textwidth]{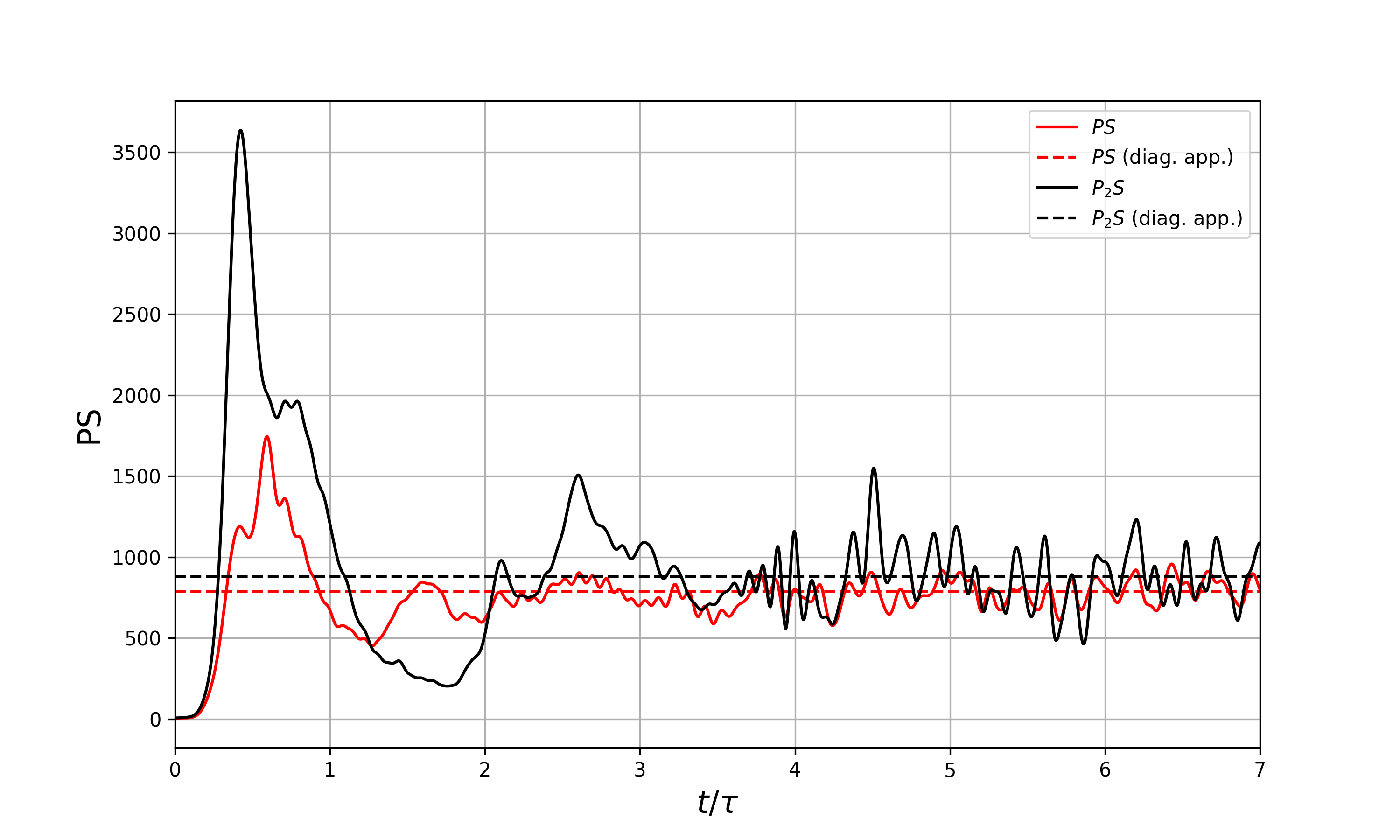}}
\caption{$PS$ vs Time for the evolution of a coherent state in position $(Q_{x}=1.3,Q_{y}=0.5)$ and momentum $(P_{x}=28,P_{y}=5)$. The time is measured in terms of the characteristic time between bounces in the stadium , $\tau=m(L_s+R)/p$  with the momentum $p=\sqrt{P^{2}_x+ P^{2}_y}$. Red and black curves correspond to $PS$ and $P_2S$ respectively where dashed lines represent the values with diagonal approximation.} .
\label{Fig:PvsTime}
\end{figure}

Coherent states are known to be the more classical-like states, that is the states with minimal position and momentum dispersion. 
Energy conservation ensures that 
\[
E=\frac{1}{2m}\left(\langle\hat{p}_{x}^{2}\rangle+\langle\hat{p}_{y}^{2}\rangle\right)=k_{B}T=\text{constant},
\]
so even if the local temperature (Eq. (\ref{eq:localTemp})) is not homogeneous inside the billiard the average temperature will remain constant during the evolution. However, this is not the case for the pressure. The initial coherent state not “touching” the boundary of the billiard implies a null pressure, while the evolution of the wave function implies “bouncing” off  the boundary and dispersion throughout the stadium and its boundary with the corresponding pressure variations. In Fig. \ref{Fig:PvsTime} we display the evolution of the pressure on the walls of the stadium for a wave function originally in a coherent state. The time is measured in terms of the characteristic time between collisions in the stadium and $\tau=m(L_s+R)/p$  where $p=\sqrt{P^{2}_x+ P^{2}_y}$ is the characteristic momentum of the wave function. We can initially observe large fluctuations up to a time when their amplitudes get smaller.

This can be explained by the non orthogonality of the normal derivatives at the boundary of the billiard. Indeed, suppose an initial state $\psi(x,y)$ that is decomposed in the basis of eigenfunctions $\varphi_{i}(x,y)$ of the billiard
\begin{equation}
\psi(x,y)=\sum_{j}c_{j}\varphi_{j}(x,y),\label{eq:WaveFunc}
\end{equation}
evolving as: 
\[
\psi_{t}(x,y)=\sum_{j}c_{j}e^{-iE_{j}t/\hbar}\varphi_{j}(x,y).
\]
The pressure in Eq. (\ref{eq:PresionQ}) as a function of time is then
\[
P(t)=\frac{\hbar^{2}}{2mL}\sum_{j,k}c_{j}^{*}c_{k}e^{i(E_{j}-E_{k})t/\hbar}\oint_{B}\frac{\partial\varphi_{j}^{*}}{\partial n}\frac{\partial\varphi_{k}}{\partial n}dl.
\]
The off-diagonal elements of the boundary integral do not vanish, that is,
\begin{equation}
P_{ij}\equiv\frac{\hbar^{2}}{2m}\frac{1}{L}\oint_{B}\frac{\partial\varphi_{i}^{*}(x,y)}{\partial n}\frac{\partial\varphi_{j}(x,y)}{\partial n}dl\neq A_{ij}\delta_{ij}\label{eq:nonorto}
\end{equation}
As such, the pressure varies in time and can be expressed as
\begin{equation}
P(t)=\sum_{i}\left|c_{i}\right|^{2}P_{i}+\sum_{j\neq k}c_{j}^{*}c_{k}e^{i(E_{j}-E_{k})t/\hbar}P_{jk}
\label{eq:PnoDiag}
\end{equation}
where $P_{i}$ stands for the pressure exerted by the eigenfunctions $\varphi_{i}(x,y)$ of the billiard and the off-diagonal terms are defined in Eq. (\ref{eq:nonorto}).
In order to test the IGL we performed a time average for the values of the pressure once the transient of large fluctuations has vanish. In this respect, we set the initial coherent states with different values for the momenta (hence, different temperatures), we evolve them and calculate the time average pressure after the transient.

\subsubsection{IGL for CS in the Rectangular Billiard}

\begin{figure}[t]
\includegraphics[width=0.47\textwidth]{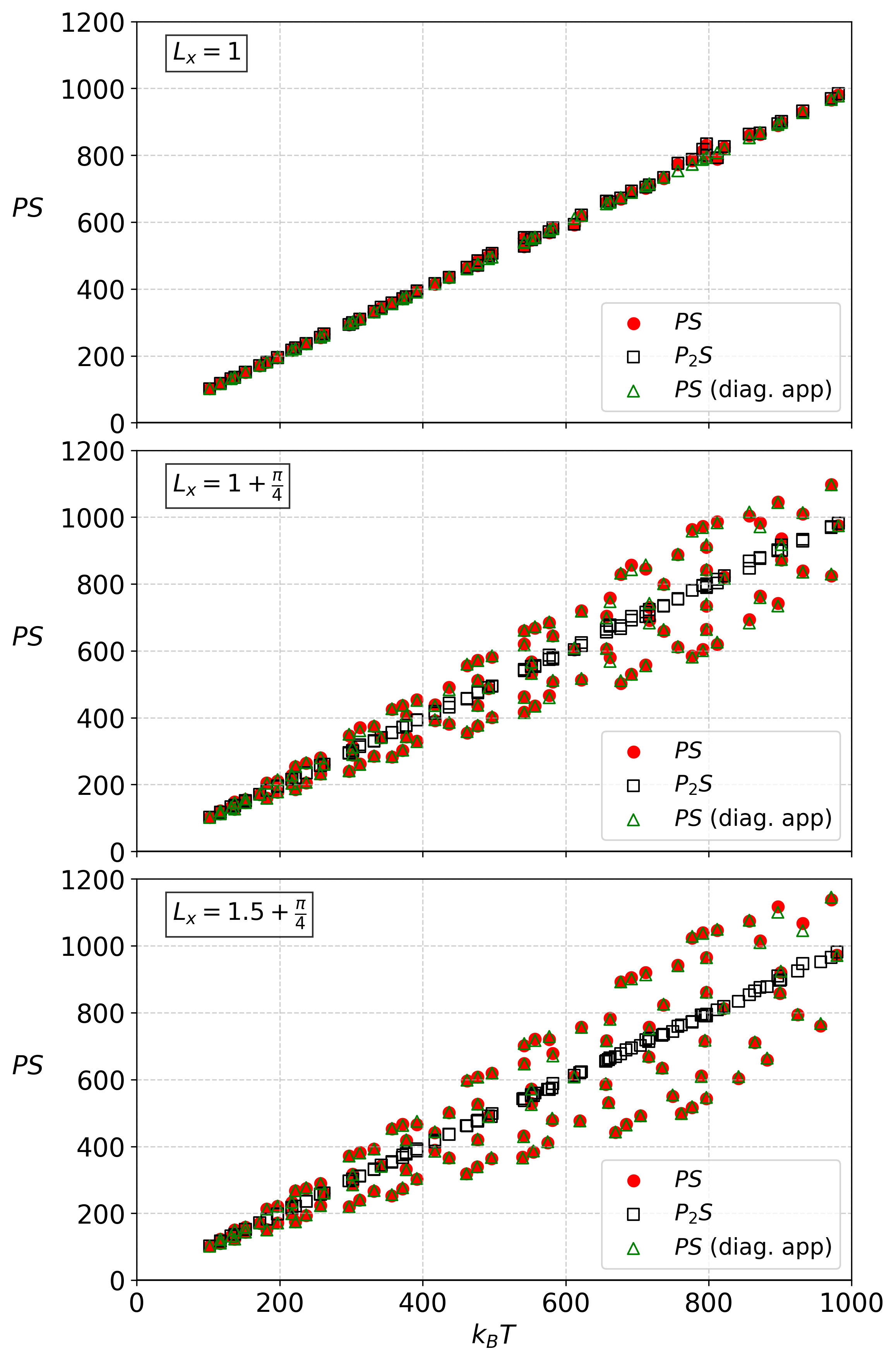}
\caption{
$PS$ versus $k_B T$ for Coherent States in the Rectangular Billiard($L_y = 1$). The panels display results for different aspect ratios: top ($L_x = 1$), middle ($L_x = 1 + \pi/4$), and bottom ($L_x = 1.5 + \pi/4$). In all panels, filled red circles correspond to $PS$, open green triangles correspond to $PS$ with the diagonal approximation Eq. (\ref{eq:Pdiag}) and open black squares correspond to $P_2S$. Initial state in coordinates is $Qx=0.2,  Qy= 0.4$ for all cases.}
\label{FIG-CS-REC}
\end{figure}

For the case of the rectangular billiard in Fig. \ref{FIG-CS-REC} we display $PS$ vs $k_{B}T$ obtained for initial coherent states of different temperatures (momenta).

It is clear that the IGL for CS in the rectangular billiard holds on average, however the  regular behavior of the eigenstates seen in Fig. \ref{FigPSvsKT-EIG-REC} is no longer present, and the dispersion has decreased. 

We also display the pressure exerted by the coherent state in the diagonal approximation where the terms involving the off-diagonal entries, $P_{ij}$ in Eq. (\ref{eq:PnoDiag}) are excluded.   The total pressure is considered as a weighted sum of the partial pressures of the eigenstates i.e.
\begin{equation}
P=\sum_{i}\left|c_{i}\right|^{2}P_{i}.
    \label{eq:Pdiag}
\end{equation}
As also be seen in the same figure, \ref{FIG-CS-REC} for the rectangular billiard, the diagonal approximation give results which are almost indistinguishable from the exact ones, suggesting that the time averaged sum of off-diagonal terms vanishes for the evolved state.

\subsubsection{IGL for CS in the Bunimovich Stadium}
\begin{figure}[t]
\includegraphics[width=0.47\textwidth]{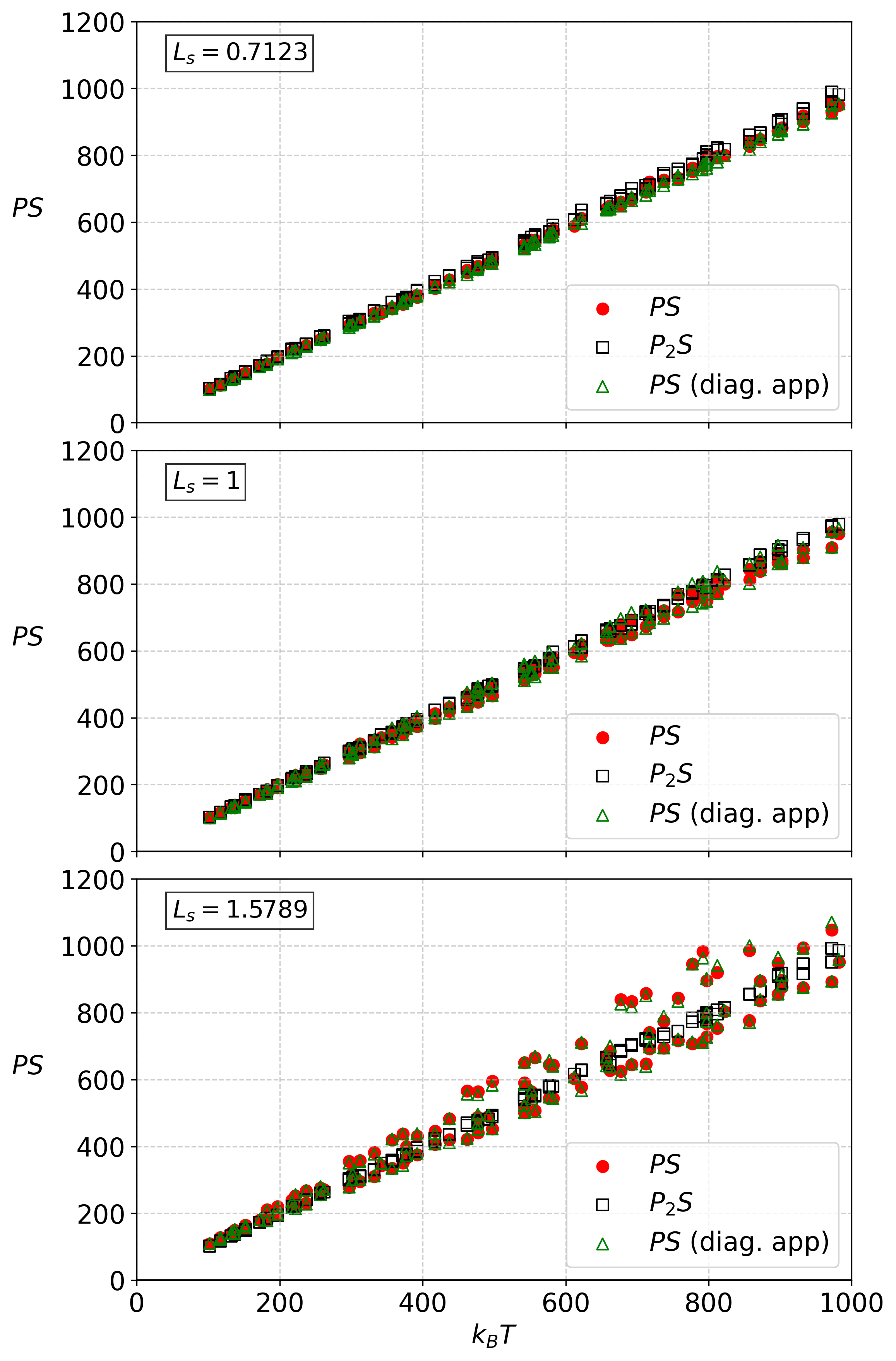}
\caption{$PS$ versus $k_B T$  for Coherent States in the Bunimovich stadium. The panels display results for different aspect ratios: top ($L_s = 0.7123$), middle ($L_s = 1$), and bottom ($L_s = 1.5789$). In all panels, filled red circles correspond to $PS$, open green triangles correspond to $PS$ with the diagonal approximation Eq. (\ref{eq:Pdiag}) and open black squares correspond to $P_2S$. Initial state in coordinates is $Qx=0.2,  Qy= 0.4$ for all cases.}
\label{FigPSvsKT-CS-Stad}
\end{figure}

In Fig. \ref{FigPSvsKT-CS-Stad} we display the results for $PS$ vs $k_{B}T$ obtained for initial coherent states in the Bunimovich stadium together with the same diagonal approximation considered previously.

It is clear that for the Bunimovich stadium the diagonal approximation gives again results which are almost indistinguishable from the exact ones.

As we show in Appendix B, the diagonal approximation is closely related to the Eigenstate Thermalization Hypothesis (ETH). Here we show that these concepts are compatible for the CS in the rectangular and Bunimovich billiards. 

\subsection{Anisotropy and dispersion}

Isotropy has revealed itself as the crucial property in order to recover the IGL behavior, we here  provide with a more quantitative treatment. We introduce the Anisotropy Index (AI) as 
\[
AI= 1 - \frac{I_-}{I_+}\,
\]
where $I_-$ and $I_+$ stand for lower and higher moments of inertia of a given bi-dimensional domain, respectively.
For example, in the case of a \textbf{rectangle} of sides' lengths \(L_x\) and \(L_y\) (assuming \(L_x \ge L_y\)), the anisotropy index is
\[
AI_{\text{rectangle}} = 1 - \frac{L_y^2}{L_x^2}\,.
\]
For a \textbf{stadium} (Bunimovich billiard) with a straight segment of length \(L_s\) and a semicircle of radius \(R\), the anisotropy index is 
\[
AI_{\text{stadium}} = 1 - \frac{16k^3 + 12\pi k^2 + 32k + 3\pi}{16k + 3\pi},
\]
where $k=L_s/R$. In order to characterize the deviation from the IGL we compute the Mean Relative Dispersion (MRD) defined by
\begin{equation}
    \sigma =  \sqrt{\frac{1}{N} \sum_i \frac{(P_i S-k_B T_i)^{2}}{(k_B T_i)^{2}}}
\end{equation}
where the points are denoted by the subscript $i$.

\begin{figure}[t]
    \centering
\includegraphics[width=0.47\textwidth]{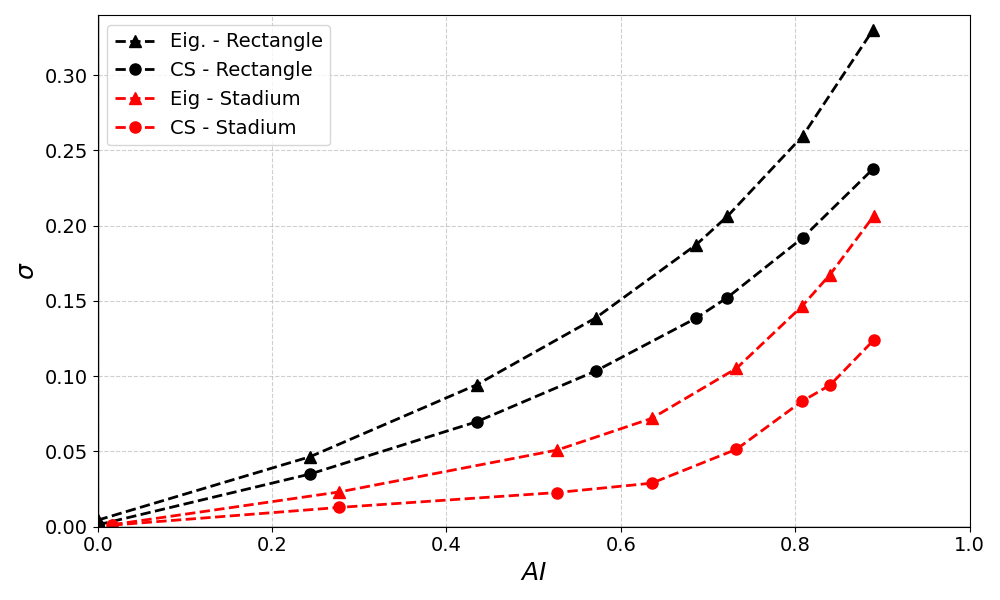}
    \caption{Dispersion from the IGL: Mean relative dispersion $\sigma$ vs the Anisotropy Index $AI$ for eigenstates and coherent sates in the rectangle and in the stadium. }
    \label{fig:Fig-sigma}
\end{figure}

In Fig. \ref{fig:Fig-sigma} we display the dispersion, characterized by $\sigma$, around the IGL for the rectangle and for the stadium billiard for both their eigenstates as well as for coherent states.

As can be seen,  in all cases, the larger the anisotropy the larger the dispersion around the IGL. In addition, for the  same value of the anisotropy index, the Bunimovich’s stadium exhibits smaller dispersion than that of the rectangle(regular), underlining the crucial role of chaos in order to recover the IGL behavior.
In both cases, for the same anisotropy the coherent states give rise to  a smaller values of  $\sigma$, which display a behavior closer to the IGL.

\section{Conclusions}

We have tested the general abiding of a single particle quantum wave function by the IGL. In this context, temperature is defined by directly extending the energy equipartition result to the quantum realm. The mean pressure is defined in two ways in order to compare a direct application of the radiation pressure concept and one that takes advantage of a quasi-orthogonality relation for billiards. In the first scenario, our results reveal that the IGL holds exactly for the circular billiard due to isotropy, while it remains valid on average for the cases of the rectangular and stadium billiard but the dispersion around the IGL increases with temperature and anisotropy. Chaotic dynamics significantly diminishes the dispersion around the IGL prescription. For coherent states,  both in the rectangular and in the Bunimovich stadium, the dispersion also lowers despite non homogeneity in temperature and pressure time fluctuations. We have observed that our diagonal approximation is equivalent to performing a pressure time average.
This approximation bears similarity to the ETH, where expectation values of observables compatible with it are indistinguishable from those calculated over the ensemble average. It is remarkable that the second definition of pressure allows us a good fit of IGL.  In summary, by linking pressure fluctuations to eigenstate statistics and dynamical chaos, our work clearly correlates the properties of quantum dynamics with the emergent thermodynamic behavior, offering insights into thermalization and the quantum-classical transition. In the future we will look for more details of this correlations and possible applications in experiments as for example measuring temperature via the pressure on the boundary.

\section*{Acknowledgement} We thank Rui Pitanga Marques da Silva and Ignacio Urrutia for their valuable suggestions. Financial support from CONICET: PIP 2022-2024 GI - 11220210100208CO is gratefully acknowledged.

\bibliographystyle{apsrev} 
\bibliography{references}

\begin{appendices}
\section{ IGL for circular billiard}

The eigenfunctions of a circular billiard (a quantum particle confined in a circular region) can be found by solving the Schrodinger equation in polar coordinates. For radius $\ensuremath{R}$, the time-independent Schrodinger equation is
\[
-\frac{\hbar^{2}}{2m}\nabla^{2}\psi(r,\theta)=E\psi(r,\theta),
\]
where $\ensuremath{\nabla^{2}}$ is the Laplacian in polar coordinates given by
\[
\nabla^{2}=\frac{1}{r}\frac{\partial}{\partial r}\left(r\frac{\partial}{\partial r}\right)+\frac{1}{r^{2}}\frac{\partial^{2}}{\partial\theta^{2}}.
\]
We assume a separable solution of the form
\[
\psi(r,\theta)=\kappa(r)\Theta(\theta).
\]
Substituting this ansatz into the Schrodinger equation and separating the variables, we get two ordinary differential equations. For the angle 
\[
\frac{d^{2}\Theta}{d\theta^{2}}+j^{2}\Theta=0,
\]
with the solution
\[
\Theta(\theta)=Ae^{ij\theta}+Be^{-ij\theta},
\]
where$\ensuremath{j}$ is an integer (due to periodic boundary conditions $\ensuremath{\Theta(\theta+2\pi)=\Theta(\theta)}).$ For the radius 
\[
r^{2}\frac{d^{2}\kappa}{dr^{2}}+r\frac{d\kappa}{dr}+\left(k^{2}r^{2}-j^{2}\right)\kappa=0,
\]
where $\ensuremath{k=\frac{\sqrt{2mE}}{\hbar}}.$ This is the Bessel differential equation, and its solution is the Bessel function of the first kind $J_{j}(kr)$. The Bessel function of the second kind $Y_{j}(kr)$ is not considered because it diverges at $r=0$. Therefore, the radial part of the wave function is:
\[
\kappa(r)=CJ_{j}(kr).
\]
The boundary condition requires that the wave function vanish at the edge of the circular billiard, $r=R$, that is $J_{j}(kR)=0$. The allowed values of $\ensuremath{k}$ are the zeros of the Bessel function $J_{j}$, denoted by $k_{jn}=\frac{x_{jn}}{R}$, where $x_{jn}$ is the $n$-th zero of $J_{j}$. Thus, the eigenfunctions of the circular billiard can be written as follows
\[
\psi_{jn}(r,\theta)=C_{jn}J_{j}\left(\frac{x_{jn}}{R}r\right)\left(Ae^{ij\theta}+Be^{-ij\theta}\right),
\]
with the corresponding eigenenergies:
\begin{equation}
E_{jn}=\frac{\hbar^{2}}{2m}\left(\frac{x_{jn}}{R}\right)^{2}=k_{B}T.\label{eq:KTQcirc}
\end{equation}
In order to calculate the pressure we need to perform the radial (normal) derivative of the wave function at the boundary $r=R$ which is 
\begin{equation}
\left.\frac{\partial\psi_{jn}}{\partial r}\right|_{r=R}=C_{jn}e^{ij\theta}\frac{x_{jn}}{R}J'_{j}(x_{jn}).
\end{equation}
Here $J'_{j}(x)$ is the derivative of the Bessel function. The mean value of the square of the normal momentum on the boundary is given by: 
\begin{equation}
\left\langle \left|\hat{p}_{r}\right|^{2}\right\rangle =\left\langle \left|-i\hbar\frac{\partial\psi_{jn}}{\partial r}\right|^{2}\right\rangle =\hbar^{2}\left\langle \left|\frac{\partial\psi_{jn}}{\partial r}\right|^{2}\right\rangle.
\end{equation}
Using the Bessel function relation 
\begin{equation}
J'_{j}(x)=\frac{j}{x}J_{j}(x)-J_{j+1}(x),
\end{equation}
and knowing that $J_{j}(x_{jn})=0$ at the boundary 
\begin{equation}
J'_{j}(x_{jn})=-J_{j+1}(x_{jn}).
\end{equation}
Substituting this into the expression for the radial derivative: 
\begin{equation}
\left.\frac{\partial\psi_{jn}}{\partial r}\right|_{r=R}=C_{jn}e^{ij\theta}\frac{x_{jn}}{R}(-J_{j+1}(x_{jn})).
\end{equation}
The mean value of the square of the normal momentum becomes
\begin{equation}
\left.\left\langle \left|\frac{\partial\psi_{jn}}{\partial r}\right|^{2}\right\rangle \right|_{r=R}=\left|C_{jn}\right|^{2}\left(\frac{x_{jn}}{R}\right)^{2}\left|J_{j+1}(x_{jn})\right|^{2}.
\end{equation}
Using the normalization constant: 
\begin{equation}
\left|C_{jn}\right|^{2}=\frac{1}{\pi R^{2}J_{j+1}^{2}(x_{jn})},
\end{equation}
we get 
\begin{equation}
\left.\left\langle \left|\frac{\partial\psi_{jn}}{\partial r}\right|^{2}\right\rangle \right|_{r=R} =\frac{x_{jn}^{2}}{\pi R^{4}J_{j+1}^{2}(x_{jn})}J_{j+1}^{2}(x_{jn}).
\end{equation}
The $J_{j+1}^{2}(x_{jn})$ terms cancel out: 
\begin{equation}
\left\langle \left|\hat{p}_{r}\right|^{2}\right\rangle =\hbar^{2}\frac{x_{jn}^{2}}{\pi R^{4}}.
\end{equation}
Then we find that the mean pressure is \[ P =\frac{1}{2m}\left\langle \left|\hat{p}_{n}\right|^{2}\right\rangle =\frac{\hbar^{2}x_{jn}^{2}}{2m\pi R^{4}}.
\]
Remembering that the average temperature is obtained as in Eq. (\ref{eq:KTQcirc}) and that $S=\pi R^{2}$ is the surface of the circular billiard, we get that for the eigenstates 
\[
PS=k_{B}T.
\]
That is, the IGL is exactly obeyed by all the eigenstates of the quantum circular billiard.

\section{Diagonal Approximation and ETH}

It is worth mentioning the deep relation between the diagonal approximation here introduced with the Eigenstate Thermalization Hypothesis
(ETH) where expectation values of observables compatible with it can be calculated either via the 
ensemble average or directly with the density matrix of equilibrium states. From the wave function (\ref{eq:WaveFunc})
we get that the density matrix is 
\[
\hat{\rho}=\sum_{i,j}c_{i}^{*}c_{j}\left|\varphi_{i}\right\rangle \left\langle \varphi_{j}\right|=\sum_{i,j}\rho_{ij}\left|\varphi_{i}\right\rangle \left\langle \varphi_{j}\right|.
\]
The temperature is obtained as 
\[
k_{B}T=\left\langle \psi\right|\frac{\hat{p}^{2}}{2m}\left|\psi\right\rangle =\frac{1}{2m}Tr\left[\hat{p}^{2}\hat{\rho}\right]=\frac{1}{2m}\sum_{i}p_{i}^{2}\rho_{ii}, 
\]
where in the last relation we can observe that only the diagonal terms
of the density matrix are involved. The pressure
can also be expressed in terms of the density matrix as
\begin{equation}
P=\frac{1}{2m}Tr\left[\Pi_{B}p_{n}\hat{\rho}p_{n}\Pi_{B}\right]=\frac{1}{2m}Tr\left[p_{n}\Pi_{B}p_{n}\hat{\rho}\right].
\label{eq:PcomoTr}
\end{equation}
 With $p_{n}$ the momentum normal to the boundary and $\Pi_{B}$
the projector on the boundary, such that 
\[
\Pi_{B}\psi(x,y)=\begin{cases}
\psi(x,y) & \text{for \ensuremath{(x,y)} on the boundary}\\
0 & \text{otherwise}
\end{cases}.
\]
The pressure is then 
\[
P=\frac{1}{2m}\sum_{i,j}\rho_{ij}Tr\left[\Pi_{B}p_{n}\left|\varphi_{i}\right\rangle \left\langle \varphi_{j}\right|p_{n}\Pi_{B}\right].
\]
Notice that in Eq. (\ref{eq:PcomoTr}) for the pressure, the operator
involved in the trace: $\hat{O}=p_{n}\Pi_{B}p_{n}$ does not commute
with the Hamiltonian $\hat{H}=\frac{\hat{p}^{2}}{2m}$, hence they
do not share the same eigenstates. This implies that  
\[
Tr\left[\hat{O}\hat{\rho}\right]\neq\sum_{i}O_{i}\rho_{ii}
\]
with $O_{i}$ the eigenvalues or equivalently, that from Eq. (\ref{eq:nonorto}) is
\begin{equation}
\frac{1}{2m}Tr\left[\Pi_{B}p_{n}\left|\varphi_{i}\right\rangle \left\langle \varphi_{j}\right|p_{n}\Pi_{B}\right]=P_{ij}\neq A_{ij}\delta_{ij}.
\label{eq:nonorto-1}
\end{equation}
Hence the diagonal approximation, consisting in eliminating the terms
involving non diagonal elements of $P_{ij}$ relates the observed pressure to the
ETH.

\end{appendices}
\end{document}